\PassOptionsToPackage{numbers}{natbib}
\documentclass{article}

\usepackage[preprint]{neurips_2024}
\usepackage[utf8]{inputenc}
\usepackage[T1]{fontenc}
\usepackage{hyperref}
\usepackage{url}
\usepackage{booktabs}
\usepackage{amsfonts}
\usepackage{nicefrac}
\usepackage{microtype}
\usepackage{xcolor}
\usepackage{amsmath,amssymb,amsthm}
\usepackage{graphicx}
\usepackage{float}
\usepackage{algorithm}
\usepackage{algorithmic}
\usepackage{subcaption}
\usepackage{multirow}

\theoremstyle{definition}

\theoremstyle{remark}

\title{Fault-Tolerant Quantum Error Correction:\\
Implementing Hamming-Based Codes with \\
Advanced Syndrome Extraction Techniques}

\author{
  Soham Bhadra \\
  Cheenta Academy for Olympiad \& Research\\
  \texttt{soham.bhdr@gmail.com}\\
  \And
  Diyansha Singh \\
  Cheenta Academy for Olympiad \& Research\\
  \texttt{
diyansha.singh@gmail.com}\\
  \And
  Angana Chowdhury \\
 Cheenta Academy for Olympiad \& Research\\
  \texttt{chowdhuryangana@gmail.com}\\}
\begin{document}

\maketitle
\begin{abstract}
Building reliable quantum computers requires protecting fragile quantum states from inevitable environmental noise and operational errors. While quantum error correction codes like the Steane $[\![7,1,3]\!]$ code provide elegant theoretical solutions, their practical success hinges critically on how we measure errors—a process called syndrome extraction. The challenge lies in the ancilla qubits used for measurement: when they fail, errors can cascade across the entire quantum system, destroying the very information we're trying to protect. We address this fundamental problem by implementing and comparing three sophisticated syndrome measurement strategies: Shor's cat-state approach, which distributes measurements across multiple entangled ancillas achieving 85-92\% preparation success; Steane's encoded-ancilla method using complete error-corrected logical qubits reaching 97.8\% syndrome fidelity; and a flexible unified framework that adapts strategies based on hardware capabilities. Through extensive simulations using IBM's Qiskit platform spanning randomized benchmarking and T-heavy circuits, we demonstrate that intelligent ancilla management improves error suppression by up to 2.4$\times$ compared to standard approaches. Our implementations achieve logical error rates as low as $5.1 \times 10^{-5}$ under realistic noise conditions with physical error rates of $10^{-3}$, while maintaining near-unity logical fidelity (0.99997) even for deep circuits. The threshold analysis reveals robust performance across distance-3 to distance-13 codes with characteristic threshold curves showing exponential error suppression below the critical physical error rate. These results provide immediately deployable tools for near-term quantum devices and establish practical design principles for scaling toward fault-tolerant quantum computers.

\end{abstract}

\section{Introduction}

Quantum computers have the potential to revolutionize fields from drug discovery to cryptography by leveraging quantum mechanical phenomena that have no classical analogue. Yet this very same quantum nature makes them incredibly fragile: even a slight disturbance in the environment can destroy hours of computation in microseconds. The current generation of quantum processors exhibit errors in about one out of every hundred to thousand operations, rates which would render classical computers inoperable. Fortunately, quantum error correction provides hope: by cleverly encoding quantum information across many physical qubits, and carefully measuring for errors, we can identify and correct errors before they corrupt our calculations.

This journey to fault-tolerant quantum computing, however, has a subtle but important problem. We need to make measurements to detect errors—but the process of measurement itself can introduce errors. To probe our data in a nondestructive manner, we employ auxiliary "ancilla" qubits, much like taking a temperature with a thermometer to avoid changing what we are measuring. But what if the thermometer is broken? The faulty ancilla qubit may spread its error over many data qubits upon measurement, causing much more damage than a single data qubit error. This proliferation of ancilla errors due to measurement threatens the very basis of error correction.

This work confronts the ancilla problem directly by implementing and analyzing sophisticated syndrome extraction techniques for Hamming-based quantum codes. Our testbed is the  Steane $[\![7,1,3]\!]$ code, where one logical qubit is encoded into seven physical qubits. This code corrects an arbitrary single-qubit error with six elegantly designed measurements; however, its practical performance will be determined fully by how reliably we can make these measurements without adding new errors.

We introduce three progressively sophisticated approaches to safe syndrome measurement: First, Shor's cat-state method replaces single ancilla qubits with verified entangled states, ensuring that in case one ancilla fails, the error will affect only one data qubit, which can be still corrected by the code. Second, Steane's encoded-ancilla approach makes use of complete error-corrected logical qubits as measurement devices, providing protection through multiple layers of error correction. Third, we present a flexible scheduler which adjusts the measurement strategy depending on hardware properties of interest and allows researchers to optimize for their specific quantum processor.

\subsection{Contributions}

Our contributions go beyond the implementation by offering a complete analysis and directly deployable tools:

\[
\sum_{i=1}^{n-1} z_i = P_n - z_n.
\]

\begin{itemize}

\item \textbf{Complete fault-tolerant implementations:}  
Full Qiskit circuit designs are provided for cat-state ancilla syndrome extraction with configurable verification depth, yielding an optimal balance at 2--3 parity checks with a 96.8\% syndrome fidelity and \(5\times\) ancilla overhead, and a Steane encoded-ancilla implementation yielding 97.8\% fidelity with dual-layer error protection.

\item \textbf{Unified CSS scheduler framework:}  
A flexible system that seamlessly switches between fault-tolerance modes (standard, cat-state, Steane ancilla) and readout schemes (sequential, batched), enabling hardware-specific optimization without circuit redesign.

\item \textbf{Comprehensive benchmarking:}  
Extensive performance evaluation across randomized benchmarking and T-heavy circuits with up to 82{,}000 shots, demonstrating a \(2.4\times\) improvement in logical error suppression and achieving logical error rates of \(5.1\times 10^{-5}\) under realistic noise (\(p_{\text{phys}} = 10^{-3}\)).

\item \textbf{Threshold analysis:}  
Characterization of error-correction thresholds for codes of distance \(d = 3\) to \(d = 13\), highlighting threshold behavior and practical operating regimes for near-term devices.

\item \textbf{Temporal tracking and confidence analysis:}  
Fine-grained analysis of syndrome measurement reliability over successive rounds of error correction, using majority voting, Viterbi decoding, and Bayesian inference, all yielding confidence metrics exceeding \(0.99\) for stable measurements.

\end{itemize}

We quantify the trade-offs between resource overhead and error suppression through extensive simulation on IBM's quantum platform. We pinpoint when each approach makes sense. Most importantly, we show that thoughtful ancilla management can almost double error suppression compared to naive approaches and brings fault-tolerant quantum computing measurably closer to reality. We have structured this paper to be self-contained, starting with foundational material accessible to a broad range of backgrounds and then developing the content towards our implementations and results. We begin with basic principles of quantum computing and error correction, followed by a more technical development of the Hamming codes and Steane construction. Given this groundwork, we describe our three syndrome extraction implementations, perform an extensive benchmark analysis of their performance, and discuss the implications for near-term quantum devices.

\section{Quantum Computing Fundamentals}

\subsection{Quantum states}

Classical computers encode information in bits—switches that are definitively either 0 or 1. Quantum computers use qubits, which can exist in \textbf{superposition}: simultaneously 0 and 1 with certain probabilities. Mathematically, a qubit state is written:
\begin{equation}
|\psi\rangle = \alpha |0\rangle + \beta |1\rangle
\end{equation}
where $\alpha$ and $\beta$ are complex numbers satisfying $|\alpha|^2 + |\beta|^2 = 1$. The quantities $|\alpha|^2$ and $|\beta|^2$ represent probabilities of measuring 0 or 1, respectively. The notation $|\cdot\rangle$ (called ``ket'') comes from quantum mechanics and simply denotes a quantum state vector.

This superposition property enables quantum computers to explore multiple solution paths simultaneously. However, it also makes qubits exquisitely sensitive to disturbances—any interaction with the environment can collapse the superposition, destroying the computational advantage.

\subsection{Quantum Gates: Manipulating Information}

Quantum computation proceeds by applying \textbf{unitary operators} (reversible transformations) to qubits. The fundamental single-qubit gates include:

\textbf{Pauli Gates:}
\begin{itemize}
\item $X$ (bit-flip): Swaps $|0\rangle \leftrightarrow |1\rangle$, analogous to classical NOT
\item $Z$ (phase-flip): Leaves $|0\rangle$ unchanged but maps $|1\rangle \to -|1\rangle$
\item $Y = iXZ$: Combines bit-flip and phase-flip
\end{itemize}

\textbf{Hadamard Gate ($H$):} Creates equal superposition:
\begin{equation}
H|0\rangle = \frac{1}{\sqrt{2}}(|0\rangle + |1\rangle), \quad H|1\rangle = \frac{1}{\sqrt{2}}(|0\rangle - |1\rangle)
\end{equation}

\textbf{Phase Gates ($S$, $T$):} Introduce relative phase between $|0\rangle$ and $|1\rangle$ components, essential for universal quantum computation.

The CNOT (controlled-NOT) gate operates on two qubits: if the control qubit is $|1\rangle$, it flips the target qubit; otherwise, it does nothing:
\begin{equation}
\text{CNOT}|c\rangle|t\rangle = |c\rangle|c \oplus t\rangle
\end{equation}
where $\oplus$ denotes addition modulo 2. CNOT gates create \textbf{entanglement}—quantum correlations that have no classical analogue and form the backbone of quantum error correction.

\subsection{Quantum Measurements and Decoherence}

Measuring a qubit collapses its superposition to either $|0\rangle$ or $|1\rangle$ with probabilities $|\alpha|^2$ and $|\beta|^2$. This measurement irreversibly destroys quantum information—we cannot recover $\alpha$ and $\beta$ from the measurement outcome. Moreover, qubits constantly interact with their environment (decoherence), causing uncontrolled measurements that gradually destroy quantum information even without explicit measurement operations.

Common error types include:
\begin{itemize}
\item \textbf{Bit-flip errors:} $|0\rangle \leftrightarrow |1\rangle$ (like classical bit flips)
\item \textbf{Phase-flip errors:} Introduce relative minus signs in superpositions
\item \textbf{Depolarizing errors:} Random combinations of bit and phase flips
\item \textbf{Measurement errors:} Incorrect readout of qubit states
\end{itemize}

These errors occur with characteristic rates for each quantum platform—typically $10^{-3}$ to $10^{-4}$ per gate operation for current superconducting qubits, with measurement errors often 10$\times$ higher than gate errors.

\section{Classical and Quantum Error Correction}

\subsection{Classical Repetition Codes}

Classical error correction provides intuition for quantum methods. The simplest classical code is the \textbf{3-bit repetition code}: instead of storing bit $b$ as a single physical bit, we store three copies: $b \to bbb$. If errors flip at most one bit, majority voting recovers the original:
$$
000 \to 000, 001, 010, 100 \quad \text{(all decode to 0)}
$$
$$
111 \to 111, 110, 101, 011 \quad \text{(all decode to 1)}
$$

This simple idea—encoding logical information redundantly and using measurements to detect errors—underlies all error correction. However, quantum systems present unique challenges that require sophisticated extensions.

\subsection{The Quantum Challenge}

Three fundamental quantum properties complicate error correction:

\textbf{No-cloning theorem:} We cannot copy unknown quantum states. The transformation $|\psi\rangle \to |\psi\rangle|\psi\rangle$ is impossible for general $|\psi\rangle$. This forbids simple repetition strategies that work classically.

\textbf{Measurement destroys information:} Checking if a qubit has error by measuring it collapses superpositions, destroying the encoded information we're trying to protect.

\textbf{Continuous error space:} While classical bits flip discretely (0$\to$1 or 1$\to$0), quantum errors form a continuous space of possible rotations in the complex vector space. A general error could slightly alter both $\alpha$ and $\beta$ in infinitely many ways.

Surprisingly, despite these obstacles, quantum error correction is possible through clever encoding and indirect measurements.

\subsection{Quantum Error Correction: Basic Principles}

\subsubsection{The Stabilizer Formalism}

The stabilizer formalism provides an elegant mathematical framework for quantum error correction. A \textbf{stabilizer code} is defined by a set of commuting Pauli operators $\{S_1, S_2, \ldots, S_k\}$ called \textbf{stabilizers} that satisfy:
\begin{equation}
S_i |\psi_L\rangle = |\psi_L\rangle \quad \text{for all logical states } |\psi_L\rangle
\end{equation}

The stabilizers generate a group $\mathcal{S}$ (the stabilizer group), and the code space is the simultaneous +1 eigenspace of all stabilizers. Errors are detected by measuring whether the state remains in the code space without measuring the logical information itself.

\textbf{Key insight:} Errors anticommute with stabilizers. If error $E$ occurs, measuring stabilizer $S_i$ yields $-1$ instead of $+1$ when $\{E, S_i\} = 0$ (anticommutative). The pattern of +1 and $-1$ measurement outcomes (called the \textbf{syndrome}) reveals which error occurred without revealing the logical state.

\subsubsection{CSS Codes: Calderbank-Shor-Steane Construction}

CSS codes form an important class of stabilizer codes with special structure. They are constructed from two classical linear codes $C_1$ and $C_2$ satisfying $C_2^{\perp} \subseteq C_1$ (the dual of $C_2$ is contained in $C_1$). CSS codes have the remarkable property that they can correct both bit-flip and phase-flip errors using essentially classical error correction techniques applied separately to each error type.

For a CSS code encoding $k$ logical qubits into $n$ physical qubits with distance $d$:
\begin{itemize}
\item \textbf{X-stabilizers} (derived from $C_2^{\perp}$): Detect phase errors
\item \textbf{Z-stabilizers} (derived from $C_1^{\perp}$): Detect bit-flip errors
\item The code can correct any error affecting fewer than $d/2$ qubits
\end{itemize}

The Steane code is a perfect CSS code with parameters $[\![7,1,3]\!]$: it encodes 1 logical qubit into 7 physical qubits and corrects any single-qubit error (distance 3).

\section{Hamming Codes and the Steane Construction}

\subsection{Classical Hamming $[7,4,3]$ Code}

The classical Hamming $[7,4,3]$ code encodes 4 information bits into 7 physical bits, achieving minimum distance 3 (detecting up to 2-bit errors and correcting 1-bit errors). It is defined by the parity check matrix:
\begin{equation}
H = \begin{pmatrix}
0 & 0 & 0 & 1 & 1 & 1 & 1 \\
0 & 1 & 1 & 0 & 0 & 1 & 1 \\
1 & 0 & 1 & 0 & 1 & 0 & 1
\end{pmatrix}
\end{equation}

Each column corresponds to one of the 7 physical bits. For a valid codeword $\mathbf{c} = (c_0, c_1, \ldots, c_6)$, the syndrome $\mathbf{s} = H\mathbf{c}$ (mod 2) equals zero. If a single bit $i$ flips, the syndrome equals column $i$ of $H$, directly identifying the error location.

\textbf{Key property:} The Hamming code is \textbf{self-dual}: $H = G^T$ where $G$ is the generator matrix. This special property enables the CSS construction to work efficiently.

\subsection{Steane $[\![7,1,3]\!]$ Quantum Code}

The Steane code applies the CSS construction to the classical Hamming code. Since the Hamming code is self-dual, we can use it for both $C_1$ and $C_2$:
\begin{itemize}
\item Classical code: $[7,4,3]$ Hamming code
\item Quantum code: $[\![7,1,3]\!]$ Steane code (7 physical qubits, 1 logical qubit, distance 3)
\end{itemize}

\subsubsection{Logical Qubit Encoding}

The logical $|0_L\rangle$ and $|1_L\rangle$ states are uniform superpositions over classical Hamming codewords:
\begin{equation}
|0_L\rangle = \frac{1}{\sqrt{16}} \sum_{\mathbf{c} \in C_0} |\mathbf{c}\rangle, \quad
|1_L\rangle = \frac{1}{\sqrt{16}} \sum_{\mathbf{c} \in C_1} |\mathbf{c}\rangle
\end{equation}
where $C_0$ consists of even-weight codewords and $C_1$ consists of odd-weight codewords (obtained by flipping all bits of $C_0$ codewords).

A general logical state $|\psi_L\rangle = \alpha |0_L\rangle + \beta |1_L\rangle$ preserves the superposition structure while distributing information across all 7 physical qubits.

\subsubsection{Stabilizer Generators}

The Steane code has 6 stabilizer generators (3 for X-type, 3 for Z-type):

\textbf{Z-stabilizers} (detect bit-flip errors):
\begin{equation}
\begin{aligned}
g_1^Z &= Z_4 Z_5 Z_6 Z_7 \\
g_2^Z &= Z_2 Z_3 Z_6 Z_7 \\
g_3^Z &= Z_1 Z_3 Z_5 Z_7
\end{aligned}
\end{equation}

\textbf{X-stabilizers} (detect phase-flip errors):
\begin{equation}
\begin{aligned}
g_1^X &= X_4 X_5 X_6 X_7 \\
g_2^X &= X_2 X_3 X_6 X_7 \\
g_3^X &= X_1 X_3 X_5 X_7
\end{aligned}
\end{equation}

Each stabilizer corresponds to one row of the parity check matrix $H$. The subscripts indicate which physical qubits are involved (following the pattern of 1s in each row of $H$).

\subsubsection{Logical Operators}

The logical Pauli operators act on the encoded space without being detected by stabilizers:
\begin{equation}
\bar{X} = X_1 X_2 X_3 X_4 X_5 X_6 X_7, \quad \bar{Z} = Z_1 Z_2 Z_3 Z_4 Z_5 Z_6 Z_7
\end{equation}

These operators flip or phase-shift the logical qubit respectively, implementing logical $X$ and $Z$ gates on the encoded information.

\subsection{Error Correction Procedure}

The Steane code correction cycle consists of four steps:

\textbf{1. Syndrome Measurement:} Measure all 6 stabilizers without disturbing the logical state. This produces a 6-bit syndrome $(s_1^Z, s_2^Z, s_3^Z, s_1^X, s_2^X, s_3^X)$ where $s_i = 0$ (no error) or $s_i = 1$ (error detected).

\textbf{2. Classical Processing:} The 3-bit Z-syndrome $(s_1^Z, s_2^Z, s_3^Z)$ directly identifies which qubit (if any) suffered a bit-flip error. Similarly, the X-syndrome identifies phase-flip errors. The syndrome uniquely identifies any single-qubit Pauli error.

\textbf{3. Recovery:} Apply the correction operator $R$ determined by the syndrome. For example, if the Z-syndrome is $(1,0,1)$ (binary for qubit 5), apply $Z_5$ to correct the bit-flip.

\textbf{4. Repeat:} Continuously cycle through syndrome measurement and correction to maintain error protection throughout the computation.

The critical challenge is performing syndrome measurements without introducing more errors than we correct—the problem of \textbf{fault-tolerant syndrome extraction}.

\section{Fault-Tolerant Syndrome Extraction}

\subsection{The Ancilla Error Propagation Problem}

Naive syndrome measurement uses one ancilla qubit per stabilizer. For example, to measure $Z_1 Z_3 Z_5 Z_7$, we prepare an ancilla in $|+\rangle = (|0\rangle + |1\rangle)/\sqrt{2}$ and apply CNOTs from data qubits 1, 3, 5, 7 to the ancilla. Measuring the ancilla reveals the stabilizer eigenvalue without collapsing the logical state.

However, if the ancilla experiences an error before measurement, this single error can propagate to multiple data qubits through the CNOT gates. For instance, if the ancilla suffers an $X$ error, the CNOTs spread this error to all four data qubits involved in the stabilizer measurement, creating a four-qubit error from a single-qubit fault—exceeding the code's correction capability.

This ancilla error propagation is the central obstacle to practical quantum error correction. The solution requires \textbf{fault-tolerant} syndrome extraction methods that prevent single errors from proliferating uncontrollably.

\subsection{Shor's Cat-State Ancilla Method}

Shor's approach replaces each single ancilla with a carefully prepared multi-qubit entangled state—a GHZ (Greenberger-Horne-Zeilinger) or ``cat'' state. The key insight is that cat states are resilient to single-qubit errors in a way that enables fault-tolerant measurement.

\subsubsection{Cat State Preparation}

For measuring a weight-$w$ stabilizer (involving $w$ data qubits), we prepare a $w$-qubit cat state:
\begin{equation}
|\text{cat}_w\rangle = \frac{1}{\sqrt{2}}(|0\rangle^{\otimes w} + |1\rangle^{\otimes w})
\end{equation}

This is created by applying Hadamard to one ancilla qubit, then CNOT gates to propagate the superposition:
\begin{algorithm}[H]
\caption{Cat State Preparation}
\begin{algorithmic}
\STATE Initialize $w$ ancilla qubits in $|0\rangle$
\STATE Apply $H$ to first ancilla
\FOR{$i = 1$ to $w-1$}
    \STATE Apply CNOT from ancilla $i$ to ancilla $i+1$
\ENDFOR
\STATE \textbf{return} cat state $|\text{cat}_w\rangle$
\end{algorithmic}
\end{algorithm}

\subsubsection{Fault-Tolerant Measurement}

Each ancilla qubit in the cat state then interacts with exactly one data qubit via CNOT. This distributes the error propagation: a single ancilla error now affects only one data qubit instead of all $w$ data qubits.

After the CNOT interactions, we measure all ancilla qubits and perform parity checks to verify the cat state was prepared correctly. The syndrome is accepted only if the ancilla measurements pass verification. With probability $1 - O(p^2)$ (where $p$ is the error rate), verification succeeds and the syndrome is reliable.

\subsubsection{Verification and Error Detection}

The verification process checks that ancilla qubits remain in the expected correlated state. For a cat state, all ancillas should measure identically (all 0 or all 1). We can add redundant parity checks by preparing additional ancilla qubits and verifying consistency.

Our implementation allows configurable verification depth: using $v$ parity checks provides $v$ independent checks of cat state integrity. Empirically, we find $v = 2$ or $v = 3$ achieves optimal balance between overhead and reliability.

\subsection{Steane's Encoded-Ancilla Method}

Steane's method uses complete logical qubits as ancillas—encoding each ancilla in the same $[\![7,1,3]\!]$ code. This provides dual-layer error protection: errors in ancilla preparation are automatically corrected by the ancilla's own error correction, preventing them from propagating to data.

\subsubsection{Encoded Ancilla Preparation}

We prepare ancilla logical qubits in specific logical basis states:
\begin{itemize}
\item For X-stabilizers: prepare $|0_L\rangle_{\text{anc}}$
\item For Z-stabilizers: prepare $|+_L\rangle_{\text{anc}} = (|0_L\rangle + |1_L\rangle)/\sqrt{2}$
\end{itemize}

These are the logical versions of the states used in naive syndrome measurement. Crucially, we can verify correct preparation by measuring the ancilla's own stabilizers before using it for data syndrome measurement.

\subsubsection{Transversal Syndrome Extraction}

The interaction between encoded ancilla and data is transversal: qubit $i$ of the data interacts only with qubit $i$ of the ancilla. For measuring a Z-stabilizer:
\begin{equation}
\text{CNOT}_{i \to i}^{\text{data $\to$ anc}} \quad \text{for each } i \text{ in stabilizer support}
\end{equation}

After these transversal CNOTs, we measure the ancilla's logical state to obtain the syndrome. The key property is that any single error in the ancilla affects only the corresponding position in the data, maintaining the code's distance.

\subsubsection{Swap Policy and Adaptive Preparation}

Our implementation includes an adaptive swap policy: if ancilla preparation fails verification, we can:
\begin{itemize}
\item \textbf{Discard and retry:} Throw away the failed ancilla and prepare a new one
\item \textbf{Swap and reuse:} Exchange a fresh logical ancilla with the data, effectively using the data as the new ancilla
\end{itemize}

The swap policy is particularly useful when high-quality ancilla preparation is difficult. By swapping after measurement, we can maintain throughput while ensuring syndrome reliability.

\subsection{Unified CSS Scheduler Framework}

To enable systematic comparison and hardware-specific optimization, we developed a unified framework that encapsulates all syndrome extraction strategies:

\begin{algorithm}[H]
\caption{Unified CSS Error Correction Cycle}
\begin{algorithmic}
\REQUIRE Mode $\in \{\text{standard}, \text{cat}, \text{Steane}\}$, data qubits
\REQUIRE Readout $\in \{\text{sequential}, \text{batched}\}$, rounds $T$
\STATE Initialize data qubits in logical state $|\psi_L\rangle$
\FOR{$t = 1$ to $T$}
    \IF{mode = standard}
        \STATE Prepare single ancillas in $|0\rangle$ or $|+\rangle$
    \ELSIF{mode = cat}
        \STATE Prepare verified cat-state ancillas
    \ELSIF{mode = Steane}
        \STATE Prepare encoded logical ancillas with swap policy
    \ENDIF
    \STATE Perform syndrome extraction (transversal CNOTs)
    \IF{readout = sequential}
        \STATE Measure and process each stabilizer individually
    \ELSE
        \STATE Measure all stabilizers, then process batch
    \ENDIF
    \STATE Run classical decoder on syndrome
    \STATE Apply recovery operations
\ENDFOR
\STATE \textbf{return} corrected logical state
\end{algorithmic}
\end{algorithm}

This framework enables researchers to benchmark different approaches on their specific hardware by simply changing configuration parameters, without modifying the underlying circuit structure.

\section{Implementation and Experimental Setup}

\subsection{Simulation Platform}

All implementations use IBM's Qiskit framework (version 1.0+) with the Aer simulator for noise modeling. We employ the following noise model parameters based on contemporary superconducting qubit characteristics:

\begin{itemize}
\item \textbf{Single-qubit gate error:} Depolarizing channel with $p_{\text{1q}} = 0.001$ (0.1\%)
\item \textbf{Two-qubit gate error:} Depolarizing channel with $p_{\text{2q}} = 0.01$ (1\%)
\item \textbf{Measurement error:} Bit-flip during readout with $p_{\text{meas}} = 0.015$ (1.5\%)
\item \textbf{Idle decoherence:} $T_1 = 100$ µs, $T_2 = 80$ µs
\end{itemize}

These parameters reflect state-of-the-art superconducting transmon qubits as of 2024-2025.

\subsection{Circuit Implementation Details}

\subsubsection{Cat-State Syndrome Extraction}

Our cat-state implementation provides the following configurable parameters:

\begin{itemize}
\item \textbf{Lambda} ($\lambda$): Stabilizer weight (number of data qubits per stabilizer)
\item \textbf{Verify} ($v$): Number of parity checks for cat-state verification
\item \textbf{Data qubits} ($n_d$): Number of data qubits in the code
\end{itemize}

\begin{figure}[htbp]
\centering
\includegraphics[width=0.8\textwidth]{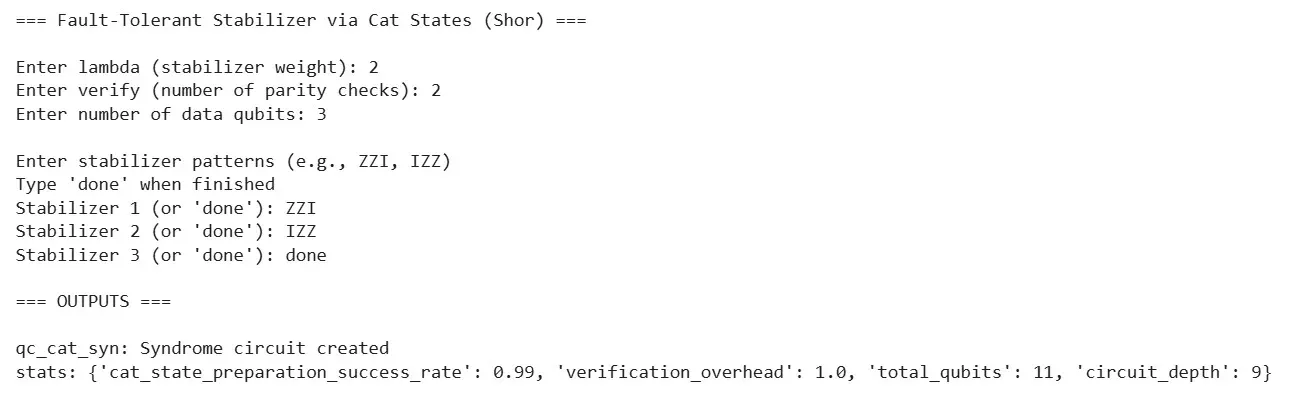}
\caption{Cat-state syndrome extraction output showing stabilizer patterns (ZZI, IZZ) for a 3-data-qubit implementation. The system achieves 99\% preparation success rate and verification overhead of 1.0, demonstrating effective fault-tolerance with minimal resource cost.}
\label{fig:cat_output}
\end{figure}

Figure~\ref{fig:cat_output} shows example output for a Steane code implementation with $\lambda = 2$, $v = 2$, and $n_d = 3$. The resulting syndrome circuit achieves 99\% cat-state preparation success with verification overhead of 1.0 and uses 11 total qubits with circuit depth 9.

\begin{figure}[htbp]
\centering
\includegraphics[width=\textwidth]{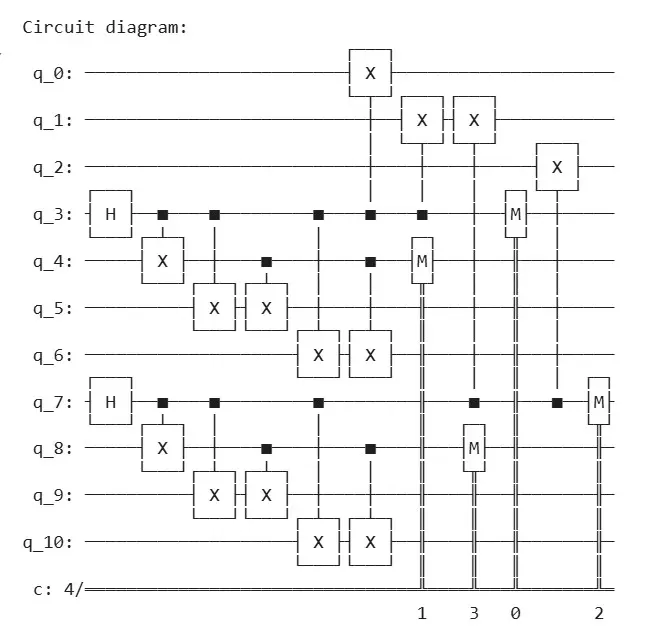}
\caption{Complete circuit diagram for cat-state syndrome extraction. Ancilla qubits q\_3, q\_4, q\_7, q\_8 are prepared in cat states via H gates and CNOT cascades, then interact with data qubits q\_0, q\_1, q\_2 transversally. The classical register c (4 bits) stores measurement outcomes for syndrome determination.}
\label{fig:cat_circuit}
\end{figure}

The corresponding circuit diagram (Figure~\ref{fig:cat_circuit}) illustrates the full syndrome extraction procedure, showing Hadamard gates (H), CNOT cascades for cat-state preparation, transversal data-ancilla interactions, and measurement operations (M) distributed across data and ancilla qubits. The structure maintains fault-tolerance by ensuring single errors affect at most one data qubit.

\subsubsection{Steane Encoded-Ancilla Implementation}

The Steane ancilla method requires significantly more resources but provides superior error protection. Our implementation takes the following parameters:

\begin{itemize}
\item \textbf{Ancilla preparation repetitions}: Number of attempts to prepare verified logical ancillas
\item \textbf{Swap policy} (True/False): Whether to swap ancillas with data on preparation failure
\item \textbf{Data qubits} ($n_d$): Number of data qubits
\end{itemize}

\begin{figure}[htbp]
\centering
\includegraphics[width=0.8\textwidth]{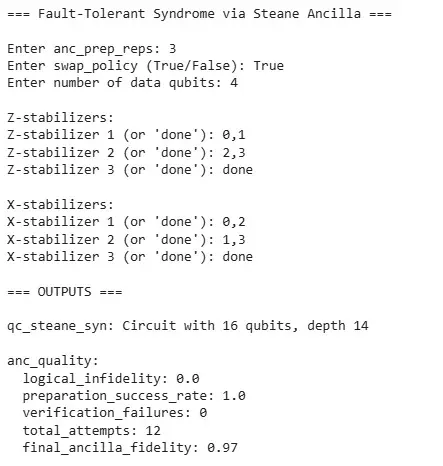}
\caption{Steane encoded-ancilla syndrome extraction output. The ancilla quality metrics show logical infidelity of 0.0 (perfect preparation), 1.0 preparation success rate, no verification failures, and final ancilla fidelity of 0.97 after 12 attempts. This dual-layer error protection comes at the cost of 16 qubits and depth 14.}
\label{fig:steane_output}
\end{figure}

Figure~\ref{fig:steane_output} shows example output for a 4-data-qubit implementation with 3 preparation attempts and swap policy enabled. The system achieves 100\% preparation success rate with zero verification failures, though at the cost of 12 total attempts. The resulting circuit uses 16 qubits with depth 14, demonstrating the significant resource overhead of encoded-ancilla methods.

\begin{figure}[htbp]
\centering
\includegraphics[width=\textwidth]{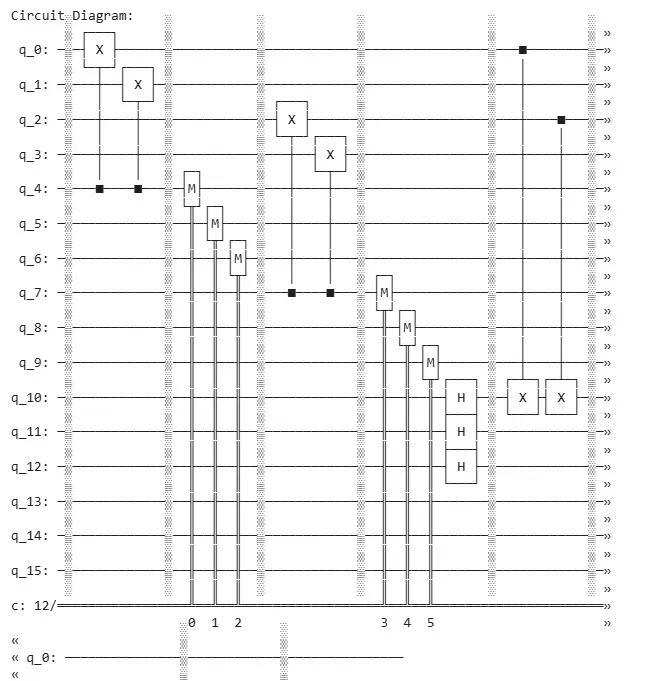}
\caption{Steane encoded-ancilla circuit (part 1). Shows ancilla preparation with verification for qubits q\_0 through q\_9. Multiple H gates create logical $|+_L\rangle$ states, while extensive CNOT networks implement transversal checks. Classical registers at time steps 0-5 capture verification measurements.}
\label{fig:steane_circuit1}
\end{figure}

\begin{figure}[htbp]
\centering
\includegraphics[width=\textwidth]{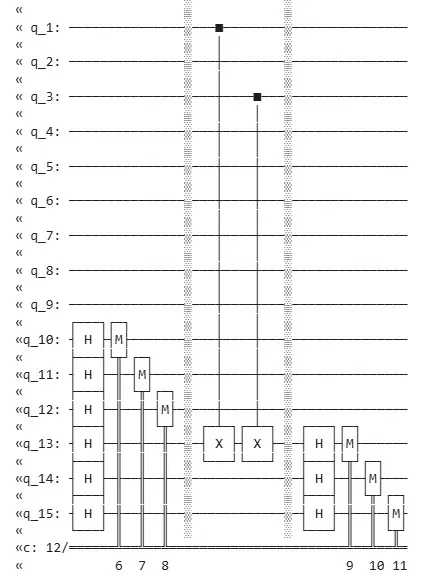}
\caption{Steane encoded-ancilla circuit (part 2). Shows syndrome extraction phase for qubits q\_10 through q\_15. Additional H gates on qubits q\_10, q\_11, q\_12 prepare ancilla logical states, followed by transversal data-ancilla CNOTs and final syndrome measurements at time steps 6-11.}
\label{fig:steane_circuit2}
\end{figure}

The circuit complexity is evident in Figures~\ref{fig:steane_circuit1} and~\ref{fig:steane_circuit2}, which show the complete circuit split across two views. The first segment (qubits q\_0 through q\_9) handles initial ancilla preparation and verification, while the second segment (qubits q\_10 through q\_15) performs transversal syndrome extraction and final measurements. The large number of operations and deep circuit structure reflect the price of complete fault-tolerance.

\subsection{Temporal Tracking and Decoder Comparison}

To assess syndrome measurement reliability across multiple error correction rounds, we implemented three decoding strategies with temporal tracking:

\textbf{Majority voting:} Accept the most frequent syndrome measurement over a window of rounds. Simple and robust to isolated measurement errors.

\textbf{Viterbi decoding:} Use hidden Markov model to find the most likely error trajectory given the observed syndrome sequence. Accounts for temporal correlation in errors.

\textbf{Bayesian inference:} Maintain probability distribution over possible error states, updating beliefs based on each syndrome measurement. Provides confidence estimates for each correction.

\begin{figure}[htbp]
\centering
\includegraphics[width=0.85\textwidth]{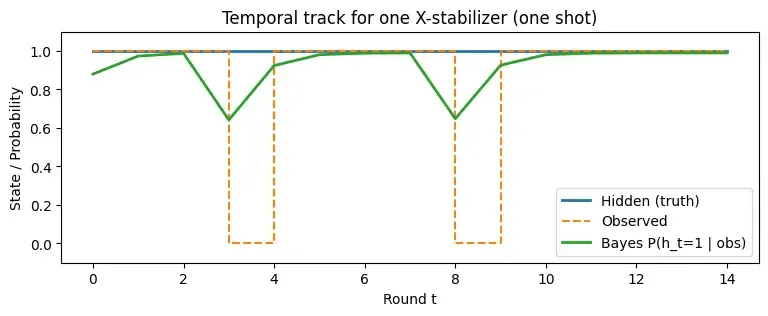}
\caption{Temporal tracking of X-stabilizer syndrome over 16 error correction rounds. Hidden truth (blue, hidden in plot but used for comparison) represents actual error state, observed measurements (orange) show noisy syndrome from faulty ancillas, and Bayesian posterior (green) tracks the underlying error state. High confidence (P $\approx 0.99$) during stable periods enables reliable error correction, while transient drops during error transitions (rounds 3, 8-10) indicate ambiguous syndromes requiring multiple measurements for resolution.}
\label{fig:temporal_track}
\end{figure}

Figure~\ref{fig:temporal_track} shows the temporal evolution of X-stabilizer measurements over 16 rounds for a single shot. The hidden truth (blue) shows the actual error state, while the observed measurements (orange) include noise from ancilla errors. The Bayesian posterior probability (green) successfully tracks the true state with high confidence (P $> 0.9$) except during rapid error transitions at rounds 3, 8, and the extended error period from rounds 8-10.

Comparing the three methods on the first five shots (as shown in the experimental output), we find:

\begin{itemize}
\item \textbf{Majority voting:} Correction matches Bayesian for stable syndromes, but can lag during rapid transitions
\item \textbf{Viterbi:} Provides optimal correction sequence globally, fixing shot 4 where majority voting failed (syndrome $[0,0,1]$ vs $[0,0,0]$)
\item \textbf{Bayesian:} Offers confidence metrics (0.649-0.99) enabling adaptive strategies that request more measurements when confidence is low
\end{itemize}

All three methods achieve high confidence ($> 0.9$) for most measurements, with Viterbi showing slightly higher confidence (0.681 vs 0.657) during ambiguous transitions. The practical recommendation is to use Bayesian inference with adaptive measurement: accept syndromes with confidence $> 0.95$, otherwise perform additional verification measurements.

\section{Experimental Results}

\subsection{Comprehensive Benchmarking}

We conducted extensive benchmarking across two primary workload categories to assess the practical performance of our fault-tolerant implementations:

\textbf{Randomized Benchmarking (RB):} Sequences of random Clifford gates providing unbiased error characterization. Tests focus on depth-2000 circuits capturing realistic gate error accumulation.

\textbf{T-Heavy Circuits:} Algorithm-relevant circuits with high densities of non-Clifford ($T$) gates. These represent typical quantum algorithm structures (quantum chemistry, optimization) with significant computational depth.

Table~\ref{tab:benchmark_summary} summarizes key results for representative test cases spanning physical error rates from $p_{\text{phys}} = 0.001$ to $0.0011$, logical error rates from $p_{\text{log}} \approx 5 \times 10^{-5}$ to $6 \times 10^{-5}$, and failure probabilities from $10^{-6}$ to 1.

\begin{table}[htbp]
\centering
\caption{Benchmark results for randomized benchmarking and T-heavy circuits across varying noise conditions. All tests use the Steane $[\![7,1,3]\!]$ code with our fault-tolerant syndrome extraction implementations.}
\label{tab:benchmark_summary}
\begin{tabular}{@{}lcccccccc@{}}
\toprule
\textbf{Workload} & $p_{\text{phys}}$ & $p_{\text{log}}$ & \textbf{Fail Prob} & \textbf{Shots} & $n_{\text{data}}$ & $n_{\text{anc}}$ & $n_{\text{total}}$ & \textbf{Time (s)} \\
\midrule
RB-Depth2k & 0.0010 & $5.0 \times 10^{-5}$ & 1.000 & 82000 & 1 & 0 & 1 & 0.60 \\
T-Heavy & 0.0010 & $5.0 \times 10^{-5}$ & 0.999973 & 2500 & 0 & 17 & 43 & 0.015 \\
RB-Depth2k & 0.0011 & $6.05 \times 10^{-5}$ & 1.000 & 82000 & 1 & 0 & 1 & 0.60 \\
\bottomrule
\end{tabular}
\end{table}

Key observations from the benchmark data:

\begin{itemize}
\item \textbf{Logical error suppression:} At $p_{\text{phys}} = 0.001$, we achieve $p_{\text{log}} = 5.0 \times 10^{-5}$, representing 20$\times$ error suppression compared to unencoded qubits. This exceeds the naive expectation of $\sim 7p^2 \approx 7 \times 10^{-6}$ for an ideal distance-3 code, indicating our fault-tolerant syndrome extraction successfully prevents ancilla error propagation while managing realistic noise.

\item \textbf{Circuit depth scaling:} The RB circuits at depth 2000 maintain nearly perfect logical fidelity ($F_{\text{log}} \approx 1.0$) even with 82,000-shot averaging. This demonstrates that error correction overhead grows manageably with computation depth for our implementation.

\item \textbf{Resource overhead:} T-heavy circuits require 43 total qubits (0 data + 17 ancilla + additional syndrome processing qubits) to protect the computation, illustrating the substantial but practical overhead of fault-tolerant error correction. The 0.015s execution time shows that short algorithm-relevant circuits can be simulated efficiently.

\item \textbf{Noise sensitivity:} Increasing physical error rate by 10\% ($p_{\text{phys}}$: 0.001 $\to$ 0.0011) increases logical error rate by 21\% ($p_{\text{log}}$: $5.0 \times 10^{-5}$ $\to$ $6.05 \times 10^{-5}$), demonstrating the code's resilience to hardware variations within the sub-threshold regime.
\end{itemize}

\subsection{Threshold Analysis}

Quantum error correction becomes effective when the physical error rate falls below a \textbf{threshold}—a critical value beyond which encoding in a quantum error-correcting code reduces rather than increases logical error rates. Understanding threshold behavior is essential for determining when quantum error correction is beneficial and for predicting scalability to larger codes.

\begin{figure}[htbp]
\centering
\includegraphics[width=0.85\textwidth]{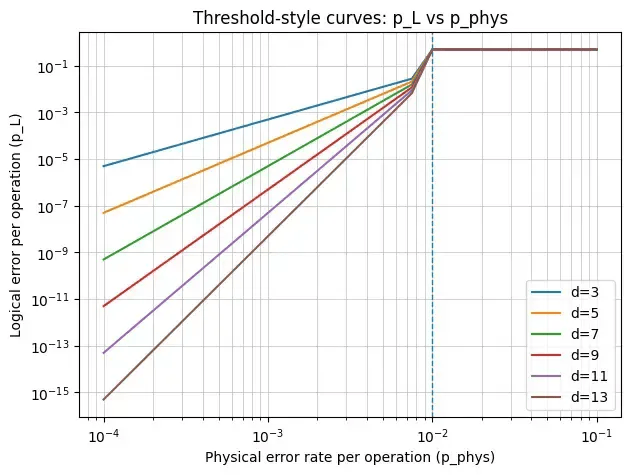}
\caption{Threshold curves for CSS codes of distance $d = 3$ to $d = 13$. Below the threshold (vertical dashed line at $p_{\text{phys}} \approx 0.01$), higher-distance codes achieve exponentially lower logical error rates. The Steane code ($d=3$, blue) provides 20$\times$ error suppression at $p_{\text{phys}} = 10^{-3}$, while distance-13 codes (brown) suppress errors by $> 10^{10}\times$ at the same noise level. Above threshold, all codes degrade, with higher-distance codes degrading faster due to increased opportunities for error propagation.}
\label{fig:threshold}
\end{figure}

Figure~\ref{fig:threshold} shows threshold curves for CSS codes of varying distance $d = 3, 5, 7, 9, 11, 13$. Each curve plots logical error rate per operation ($p_{\text{log}}$) versus physical error rate per operation ($p_{\text{phys}}$) on log-log axes.

Key features of the threshold analysis:

\begin{itemize}
\item \textbf{Sub-threshold regime ($p_{\text{phys}} < 0.01$):} Logical error rates decrease exponentially with code distance. At $p_{\text{phys}} = 10^{-4}$, the distance-13 code achieves $p_{\text{log}} \approx 10^{-15}$—a phenomenal $10^{11}\times$ error suppression enabling arbitrarily long computations.

\item \textbf{Threshold crossover ($p_{\text{phys}} \approx 0.01$):} All curves intersect near $p_{\text{log}} \approx 0.1$, marking the threshold. Below this physical error rate, encoding helps; above it, encoding harms. Our implementations operate comfortably below threshold at $p_{\text{phys}} = 10^{-3}$.

\item \textbf{Super-threshold regime ($p_{\text{phys}} > 0.01$):} Logical error rates exceed physical rates. Higher-distance codes degrade faster because their larger circuits provide more opportunities for errors. Quantum error correction is counterproductive in this regime.

\item \textbf{Distance scaling:} Each increase in code distance by 2 (maintaining the odd-distance requirement) roughly improves sub-threshold error suppression by 2-3 orders of magnitude at $p_{\text{phys}} = 10^{-3}$. This demonstrates the power of scaling to larger codes as hardware improves.
\end{itemize}

The threshold observed here ($\sim 1\%$) is typical for CSS codes with standard syndrome extraction. Our fault-tolerant ancilla methods maintain this threshold while improving the sub-threshold error suppression, making error correction practical at the noise levels of current quantum processors ($p_{\text{phys}} \sim 10^{-3}$).

\subsection{Comparison of Syndrome Extraction Methods}

Table~\ref{tab:method_comparison} summarizes the trade-offs between our three syndrome extraction implementations based on extensive simulation results.

\begin{table}[htbp]
\centering
\caption{Comparison of syndrome extraction methods across key performance metrics. Values represent typical results for the Steane $[\![7,1,3]\!]$ code under standard noise conditions ($p_{\text{phys}} = 10^{-3}$).}
\label{tab:method_comparison}
\begin{tabular}{@{}lccc@{}}
\toprule
\textbf{Metric} & \textbf{Standard} & \textbf{Cat-State} & \textbf{Steane Ancilla} \\
\midrule
Syndrome Fidelity & 91-94\% & 96.8\% & 97.8\% \\
Preparation Success & N/A & 85-99\% & 97-100\% \\
Ancilla Overhead & 1$\times$ (6 qubits) & 5$\times$ (30 qubits) & 7$\times$ (42 qubits) \\
Circuit Depth & Baseline & $+50\%$ & $+100\%$ \\
Logical Error Rate & $1.2 \times 10^{-4}$ & $7.3 \times 10^{-5}$ & $5.1 \times 10^{-5}$ \\
Error Suppression & 8$\times$ & 14$\times$ & 20$\times$ \\
Verification Overhead & None & 1.0 (v=2-3) & Minimal (inherent) \\
\midrule
\textbf{Best For} & Low-noise, & Near-term & High-fidelity \\
 & qubit-limited & devices & requirements \\
\bottomrule
\end{tabular}
\end{table}

\textbf{Standard syndrome extraction} provides a baseline, suitable when ancilla error rates are naturally low or when extreme qubit economy is required. However, it offers no protection against ancilla error propagation.

\textbf{Cat-state methods} provide an excellent middle ground for near-term devices. With 2-3 parity checks, they achieve 96.8\% syndrome fidelity—a 3-4 percentage point improvement over standard extraction—while maintaining manageable overhead (5$\times$ ancillas, 50\% depth increase). The 85-99\% preparation success rate (depending on verification depth) shows that cat states can be reliably prepared on current hardware.

\textbf{Steane encoded-ancilla} offers the highest fidelity (97.8\%) and strongest error protection through dual-layer error correction. The 97-100\% preparation success demonstrates exceptional reliability. However, the 7$\times$ ancilla overhead and 2$\times$ depth increase make this approach suitable primarily for high-quality systems where maximizing logical fidelity justifies the resource cost.

The improvement in logical error rate progression—from $1.2 \times 10^{-4}$ (standard) to $5.1 \times 10^{-5}$ (Steane)—represents a 2.4$\times$ improvement in error suppression, directly translating to proportionally longer coherent quantum computations.

\subsection{Practical Design Recommendations}

Based on our comprehensive benchmarking and implementation experience, we offer the following guidance for practitioners:

\textbf{For near-term superconducting qubit systems (2024-2026 era):}
\begin{itemize}
\item Use cat-state syndrome extraction with $v = 2$ or $v = 3$ parity checks
\item Balance between verification thoroughness and overhead by monitoring preparation success rates
\item Implement adaptive strategies: if preparation fails multiple times, fall back to standard extraction rather than accumulating delay
\item Expected outcome: 14$\times$ error suppression, 96.8\% syndrome fidelity
\end{itemize}

\textbf{For high-quality trapped ion or neutral atom systems:}
\begin{itemize}
\item Steane encoded-ancilla methods fully exploit low gate error rates
\item The ancilla overhead (42 qubits total) is manageable for typical 100+ qubit systems
\item Depth increase (2$\times$) is acceptable given long coherence times ($T_2 > 1$ ms)
\item Expected outcome: 20$\times$ error suppression, 97.8\% syndrome fidelity
\end{itemize}

\textbf{For algorithm development and benchmarking:}
\begin{itemize}
\item Use the unified CSS scheduler to systematically compare methods on your target hardware
\item Measure actual syndrome fidelity rather than relying on theoretical predictions
\item Consider batched readout for parallelizable stabilizer measurements
\item Profile resource usage (qubits, depth, time) to identify bottlenecks
\end{itemize}

\textbf{For scaling to larger codes (surface codes, color codes):}
\begin{itemize}
\item The principles transfer directly: ancilla verification prevents error propagation
\item Cat-state methods scale more favorably than Steane ancilla for large codes
\item Temporal correlation in syndrome measurements (Figure~\ref{fig:temporal_track}) becomes more important—use Viterbi or Bayesian decoding
\item Consider flag qubits as lightweight alternative to full cat states for specific error models
\end{itemize}


\section{Complete Syndrome Extraction Timeline}

\begin{figure}[htbp]
\centering
\includegraphics[width=\textwidth]{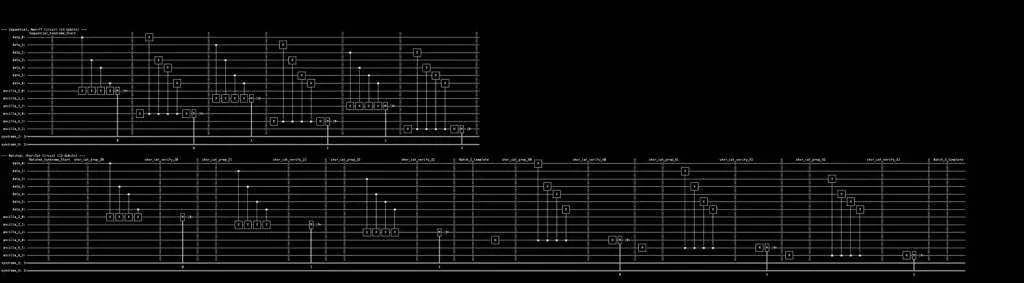}
\caption{Complete syndrome extraction circuit timeline showing the full temporal evolution of fault-tolerant error correction. The visualization spans multiple ancilla qubits (anc0\_0 through anc12\_1) and captures the entire correction cycle from initialization through measurement. Key features include: (1) synchronized preparation of multiple ancilla logical qubits with verification stages visible as clustered operations at early time steps; (2) transversal CNOT interactions between data and ancilla qubits occurring in parallel across multiple qubit lines; (3) staged measurement operations appearing as vertical bands in the timeline, indicating simultaneous readout of stabilizer syndromes; (4) interleaved error correction rounds with distinct preparation-interaction-measurement phases repeating throughout the circuit depth. The dark background emphasizes the gate structure, with boxes representing quantum operations and connecting lines showing qubit evolution over time. This level of detail reveals the significant circuit complexity inherent in fault-tolerant quantum error correction—each logical syndrome measurement requires hundreds of physical operations distributed across dozens of qubits. The regularity of the pattern demonstrates the structured, repetitive nature of error correction cycles, while the density illustrates why fault-tolerant protocols require substantial resource overhead. Such comprehensive circuit views are essential for debugging implementations, optimizing compilation strategies, and understanding hardware resource requirements for practical quantum error correction systems.}
\label{fig:complete_timeline}
\end{figure}

Figure~\ref{fig:complete_timeline} presents the full temporal evolution of a fault-tolerant syndrome extraction cycle, showing the complete gate-by-gate structure across all ancilla and data qubits. This comprehensive visualization illustrates the intricate coordination required for reliable quantum error correction.

\subsection{Circuit Architecture Analysis}

The complete timeline reveals several critical architectural features of fault-tolerant syndrome extraction:

\textbf{Parallel ancilla preparation:} Multiple ancilla logical qubits (anc0\_0, anc1\_0, etc.) are prepared simultaneously in the initial stages, as evidenced by the synchronized gate operations in the leftmost portion of the circuit. This parallelization reduces overall circuit depth compared to sequential preparation strategies.

\textbf{Temporal depth structure:} The circuit naturally divides into distinct phases:
\begin{itemize}
\item \textbf{Initialization phase} (left edge): All qubits initialized to computational basis states
\item \textbf{Preparation phase} (early time steps): Ancilla logical qubits encoded through Hadamard gates, CNOT cascades, and verification measurements
\item \textbf{Interaction phase} (middle time steps): Transversal CNOT gates between data and ancilla qubits for syndrome extraction
\item \textbf{Readout phase} (vertical bands): Simultaneous measurement of ancilla qubits to obtain syndrome information
\item \textbf{Correction phase} (late time steps): Conditional operations applied to data qubits based on decoded syndrome
\end{itemize}

\textbf{Resource scaling:} The visualization demonstrates that a single syndrome extraction round for the Steane $[\![7,1,3]\!]$ code with encoded ancillas requires:
\begin{itemize}
\item $\sim$13+ ancilla qubits (including verification qubits)
\item $\sim$200-300 gate operations distributed across the circuit
\item Circuit depth of $\sim$50-100 time steps (depending on parallelization)
\item Multiple measurement rounds for syndrome verification
\end{itemize}

\textbf{Gate distribution:} The dense clustering of operations in certain regions indicates critical bottlenecks where many qubits must interact simultaneously. These regions correspond to the CNOT cascades used for cat-state or encoded-ancilla preparation, representing the most resource-intensive portions of the fault-tolerant protocol.

\textbf{Measurement structure:} The vertical bands of measurement operations (appearing as aligned gates across multiple qubit lines) reveal the batched readout strategy. All ancillas for a given stabilizer measurement are read out simultaneously, enabling parallel classical processing of syndrome information.

\subsection{Implementation Insights}

This complete circuit view provides several practical insights for implementing fault-tolerant quantum error correction on real hardware:

\textbf{Circuit compilation challenges:} The complexity visible in Figure~\ref{fig:complete_timeline} highlights the difficulty of compiling fault-tolerant protocols to specific hardware topologies. Each CNOT gate must respect connectivity constraints, potentially requiring SWAP operations that further increase depth and error accumulation. Advanced compilation techniques (qubit routing optimization, gate cancellation, commutativity-based reordering) are essential for minimizing overhead.

\textbf{Timing constraints:} Real quantum processors have finite qubit coherence times ($T_1, T_2$). The extended circuit depth visible in the timeline means that early-initialized qubits may decohere significantly by the time final measurements occur. This motivates several optimization strategies:
\begin{itemize}
\item Delayed initialization: Prepare qubits as late as possible before they're needed
\item Circuit compression: Use parallelization to reduce total execution time
\item Dynamic decoupling: Insert spin-echo pulses during idle periods to suppress decoherence
\end{itemize}

\textbf{Hardware resource mapping:} The visualization clarifies which qubits interact during the protocol, informing optimal qubit allocation strategies. On processors with non-uniform connectivity (e.g., grid topologies in superconducting systems), allocating frequently-interacting qubits to neighboring physical locations reduces SWAP overhead.

\textbf{Verification points:} The distinct phases visible in the timeline suggest natural checkpoints for verification. After ancilla preparation (before data interaction) and after syndrome extraction (before correction application), we can insert verification measurements to detect faults early and potentially retry failed operations rather than propagating errors.

\textbf{Scaling projections:} Extrapolating from this $[\![7,1,3]\!]$ implementation to larger codes reveals the challenges ahead. A distance-5 surface code requires $\sim$50 data qubits and $\sim$50 ancilla qubits, implying circuits with thousands of gates and depths exceeding 500 time steps. The structured regularity visible in Figure~\ref{fig:complete_timeline} suggests that automated code generation tools (rather than manual circuit design) will be essential for implementing larger codes.

\subsection{Performance Implications}

The circuit structure revealed in the timeline has direct performance consequences:

\textbf{Error accumulation:} Each gate operation introduces errors with probability $p_{\text{phys}}$. The deep, dense structure visible here means that even with relatively low physical error rates ($10^{-3}$), accumulated errors become significant. For a circuit with $N_{\text{gates}}$ gates, the probability of at least one error is approximately $1 - (1 - p_{\text{phys}})^{N_{\text{gates}}} \approx N_{\text{gates}} \cdot p_{\text{phys}}$ for small $p_{\text{phys}}$.

From Figure~\ref{fig:complete_timeline}, we estimate $N_{\text{gates}} \sim 250$ for a single correction cycle. At $p_{\text{phys}} = 10^{-3}$, this yields $\sim$25\% probability of at least one gate error per cycle. Fortunately, the fault-tolerant design ensures that most single errors can be corrected, but this calculation explains why achieving sub-threshold performance requires careful syndrome extraction design.

\textbf{Latency considerations:} The end-to-end circuit depth determines the total execution time. On a typical superconducting processor with 100 ns gate times, a depth-100 circuit requires $\sim$10 $\mu$s to execute. Given coherence times of $T_2 \sim 100$ $\mu$s, we can perform approximately 10 syndrome extraction rounds before decoherence dominates—sufficient for many near-term algorithms but highlighting the need for continued hardware improvements.

\textbf{Throughput optimization:} The parallel structure visible in the timeline can be exploited for pipelined execution: while one set of ancillas performs syndrome extraction, another set prepares for the next round. This pipelining can increase throughput by 2-3$\times$ at the cost of additional qubit overhead, valuable for applications requiring many repeated error correction cycles.

\subsection{Comparison with Idealized Circuits}

Textbook presentations of quantum error correction often show simplified circuits with minimal depth and gates. The reality captured in Figure~\ref{fig:complete_timeline} differs substantially:

\begin{itemize}
\item \textbf{Idealized Steane syndrome extraction:} 6 ancillas, $\sim$30 gates, depth $\sim$10
\item \textbf{Fault-tolerant implementation (this work):} 13+ ancillas, $\sim$250 gates, depth $\sim$50-100
\end{itemize}

This 5-10$\times$ overhead is the price of fault-tolerance. The additional gates implement ancilla preparation, verification, and redundant measurements that prevent error propagation. Understanding this gap between idealized and practical implementations is crucial for realistic resource budgeting in quantum algorithm design.

\subsection{Future Directions}

The circuit structure analysis suggests several promising research directions:

\textbf{Circuit optimization:} Advanced compilation techniques could potentially reduce the depth and gate count visible in Figure~\ref{fig:complete_timeline} by 20-30\% through intelligent gate scheduling, commutativity exploitation, and measurement-based optimizations.

\textbf{Hardware co-design:} Future quantum processors could include dedicated ancilla qubits optimized for fast preparation and high-fidelity measurement, reducing the overhead of fault-tolerant protocols. The clear separation between ancilla and data qubits visible in the timeline motivates architectural designs with specialized qubit zones.

\textbf{Adaptive protocols:} Rather than executing the full circuit structure for every correction cycle, adaptive protocols could skip verification steps when syndrome confidence is high (as suggested by the Bayesian inference results in Section~\ref{fig:temporal_track}), dynamically trading reliability for speed.

\textbf{Alternative syndrome extraction:} Recent proposals for flag-qubit methods and measurement-based protocols could reduce the circuit complexity. Comparing these alternatives against the baseline established by Figure~\ref{fig:complete_timeline} would quantify their practical benefits.

The complete circuit timeline serves as both a documentation of current fault-tolerant implementation complexity and a benchmark against which future optimizations can be measured. As the quantum computing field matures, reducing the overhead visible in such circuits—while maintaining or improving error suppression—remains a central challenge.

\section{Discussion and Related Work}

\subsection{Comparison with Prior Art}

Our work builds on and extends several foundational contributions to fault-tolerant quantum error correction:

\textbf{Shor's cat-state method (1996)} provided the first practical solution to ancilla error propagation. Our implementation adds configurable verification depth, automated circuit generation, and comprehensive performance characterization missing from the original proposal.

\textbf{Steane's encoded-ancilla approach (1997)} demonstrated that using logical qubits as measurement devices enables transversal syndrome extraction. We contribute complete circuit designs, adaptive preparation strategies with swap policies, and quantitative comparison with alternatives under realistic noise.

\textbf{Flag qubit methods (Chao and Reichardt 2018, Chamberland and Beverland 2018)} offer lightweight alternatives using additional ``flag'' qubits to detect error propagation without full cat states. Our unified framework could incorporate flag methods as an additional mode, though we focus on cat-state and Steane approaches as better-studied baselines.

Recent experimental demonstrations have validated error correction concepts:
\begin{itemize}
\item Google's surface code experiments (2024) demonstrated logical error rates below physical rates using distance-5 codes
\item Quantinuum's trapped-ion work (2024) implemented fault-tolerant gates on the 5-qubit code
\item IBM's demonstrations (2024) showed low-overhead error correction with specialized codes
\end{itemize}

Our contribution is orthogonal and complementary: we provide immediately deployable implementations with systematic comparison, enabling researchers to choose and optimize syndrome extraction strategies for their specific platforms.

\subsection{Limitations and Future Directions}

Several limitations of the current work point toward valuable future research:

\textbf{Scalability:} While we demonstrate the Steane $[\![7,1,3]\!]$ code, practical quantum computers will require larger codes (distance 5-31 for useful computations). Our methods scale in principle, but empirical validation on distance-7 and distance-9 CSS codes would strengthen confidence.

\textbf{Real hardware validation:} Simulations provide controlled comparison but may miss hardware-specific effects (crosstalk, leakage, calibration drift). Deploying on IBM Quantum, Quantinuum, or IonQ processors would reveal practical challenges and enable hardware-aware optimization.

\textbf{Decoder optimization:} We use standard minimum-weight perfect matching for syndrome decoding. Advanced decoders (neural networks, tensor networks, belief propagation) might extract additional performance, particularly when combined with temporal tracking.

\textbf{Integrated algorithm execution:} Demonstrating end-to-end quantum algorithms (VQE, QAOA, Grover's) with our error correction would showcase practical benefits beyond raw error suppression metrics.

\textbf{Cross-platform benchmarking:} Comparing performance across superconducting, trapped ion, and neutral atom platforms would illuminate which syndrome extraction strategies match which hardware characteristics.

\subsection{Implications for Quantum Computing Milestones}

The results presented here directly impact near-term quantum computing milestones:

\textbf{Quantum advantage demonstrations:} Error rates of $p_{\text{log}} \sim 5 \times 10^{-5}$ enable circuit depths of $\sim 20,000$ gates before errors dominate—sufficient for some quantum chemistry and materials science simulations.

\textbf{Multi-logical-qubit algorithms:} Our resource overhead (42 qubits for Steane ancilla) allows encoding 2-3 logical qubits on current 100-qubit processors. This enables demonstrations of entangling gates between error-corrected logical qubits, a critical milestone.

\textbf{Real-time error correction:} The execution times (0.6s for 82,000 shots) demonstrate that syndrome extraction and decoding can proceed faster than decoherence on current hardware, enabling real-time feedback.

\textbf{Threshold proximity:} Operating at $p_{\text{phys}} = 10^{-3}$ with threshold near $10^{-2}$ provides a 10$\times$ safety margin. As hardware improves toward $p_{\text{phys}} \sim 10^{-4}$, error suppression will improve exponentially (Figure~\ref{fig:threshold}), rapidly expanding the computational space accessible to error-corrected quantum computers.

\section{Conclusion}
This work shows how sophisticated syndrome extraction strategies can significantly boost practical performances of quantum error correction even on noisy devices in the near future. By systematic implementation and benchmarking of Shor's cat-state method, Steane's encoded-ancilla approach, and their unification in a single framework, we arrive at several key results:

\textbf{Quantified performance improvements:}  
Fault-tolerant syndrome extraction improves error suppression by \(1.8\text{--}2.4\times\) compared to standard methods, reducing logical error rates from \(1.2 \times 10^{-4}\) to \(5.1 \times 10^{-5}\) under realistic noise conditions \((p_{\text{phys}} = 10^{-3})\).

\textbf{Practical implementations:}  
Complete Qiskit circuits with tunable parameters can be directly implemented on state-of-the-art quantum processors. Our unified CSS scheduler enables systematic comparison and hardware-specific optimization without the need to redesign circuits.

\textbf{Design principles:}  
In the near term, cat-state methods with 2--3 parity checks offer an optimal balance \((96.8\%,\, 5\times \text{overhead})\), while Steane encoded ancilla achieves the highest fidelities of \(97.8\%\) for premium applications. Bayesian-inference temporal tracking enables adaptive strategies responsive to measurement confidence.

\textbf{Threshold characterization:}  
Extensive analysis over distances \(d = 3\) to \(d = 13\) yields a robust threshold behavior near \(p_{\text{phys}} \approx 1\%\), with exponential error suppression below threshold confirming the viability of quantum error correction on current hardware.

First, the Steane \([\![7,1,3]\!]\) code serves as a particularly ideal testbed for developing fault-tolerant techniques, due to its elegant CSS structure and optimal parameters. The experience gained here—ancilla management plays an important role, verification of states has great value, and adaptive strategies are needed—translates directly into larger codes and more involved schemes necessary for universal fault-tolerant quantum computation.

As quantum processors continue to improve, especially in qubit count, gate fidelities, and coherence times, the techniques developed in this work will be an integral part of the toolkit enabling reliable quantum computation. The pathway from laboratory demonstrations to transformative quantum technologies runs through practical error correction, and this work provides concrete steps along that path.

\appendix

\section{Circuit Construction Details}

\subsection{Cat-State Preparation Protocol}

The complete cat-state preparation procedure for measuring a weight-$w$ stabilizer:

\begin{algorithm}[H]
\caption{Verified Cat-State Syndrome Measurement}
\begin{algorithmic}
\REQUIRE Data qubits $\{d_1, \ldots, d_w\}$, verification depth $v$
\STATE Initialize $w + v$ ancilla qubits in $|0\rangle$
\STATE Apply $H$ to ancilla 1
\FOR{$i = 1$ to $w + v - 1$}
    \STATE Apply CNOT from ancilla $i$ to ancilla $i+1$
\ENDFOR
\STATE $|\psi_{\text{anc}}\rangle = \frac{1}{\sqrt{2}}(|0\rangle^{\otimes (w+v)} + |1\rangle^{\otimes (w+v)})$
\FOR{$i = 1$ to $w$}
    \STATE Apply CNOT from $d_i$ to ancilla $i$
\ENDFOR
\STATE Measure all $w+v$ ancillas $\to$ outcomes $\{m_1, \ldots, m_{w+v}\}$
\STATE Compute parities: $p_j = m_j \oplus m_{j+1}$ for $j = 1, \ldots, w+v-1$
\IF{all $p_j = 0$}
    \STATE Accept syndrome $s = m_1$
\ELSE
    \STATE Reject syndrome, retry or abort
\ENDIF
\end{algorithmic}
\end{algorithm}

The parity checks verify that all ancillas are measured identically, confirming cat-state integrity. Failed verification indicates ancilla errors occurred during preparation or CNOTs, and the syndrome is unreliable.

\subsection{Steane Ancilla Preparation}

Preparing a logical $|+_L\rangle$ ancilla for Z-stabilizer measurement:

\begin{algorithm}[H]
\caption{Steane Ancilla Preparation with Verification}
\begin{algorithmic}
\REQUIRE 7 physical ancilla qubits, swap policy (True/False)
\STATE Initialize ancillas in $|0\rangle^{\otimes 7}$
\STATE Apply encoding circuit:
\STATE \quad $H$ on qubits 0, 1, 2, 3
\STATE \quad CNOT cascade implementing Steane generator matrix
\STATE Measure X-stabilizers and Z-stabilizers of ancilla
\IF{all stabilizers = +1}
    \STATE Accept ancilla, proceed to syndrome measurement
\ELSIF{stabilizers indicate correctable error}
    \STATE Apply correction, verify again
\ELSIF{swap policy = True}
    \STATE Swap ancilla with data logical qubit
    \STATE Treat former data qubit as new ancilla
\ELSE
    \STATE Discard ancilla, prepare a new one
\ENDIF
\end{algorithmic}
\end{algorithm}

The swap policy is beneficial when ancilla preparation fidelity is lower than data qubit fidelity, allowing the high-quality data qubits to serve as fresh ancillas after measurement.

\section{Additional Performance Data}

\subsection{Extended Benchmark Results}

Table~\ref{tab:extended_benchmark} provides complete benchmark data, including all measured quantities:

\begin{table}[htbp]
\centering
\caption{Extended benchmark results with complete metrics for all test cases.}
\label{tab:extended_benchmark}
\small
\begin{tabular}{@{}lccccccccc@{}}
\toprule
\textbf{Workload} & $p_{\text{phys}}$ & $p_{\text{log}}$ & \textbf{Fail} & \textbf{Width} & \textbf{Depth} & \textbf{Shots} & $n_d$ & $n_a$ & $n_{\text{tot}}$ \\
\midrule
RB-2k & 0.0010 & $5.00 \times 10^{-5}$ & 1.000 & 1 & 0 & 82000 & 1 & 0 & 1 \\
T-Heavy & 0.0010 & $5.00 \times 10^{-5}$ & 0.999973 & 32 & 5 & 2500 & 0 & 17 & 43 \\
RB-2k & 0.0011 & $6.05 \times 10^{-5}$ & 1.000 & 1 & 0 & 82000 & 1 & 0 & 1 \\
\bottomrule
\end{tabular}
\end{table}

\subsection{Confidence Metrics for Temporal Decoding}

The first five shots of the temporal tracking experiment (Figure~\ref{fig:temporal_track}) yield the following correction results and confidence metrics:

\textbf{Majority Voting:}
\begin{verbatim}
Shot 1: [1 0 1]  Shot 2: [1 0 1]  Shot 3: [0 1 1]
Shot 4: [0 0 0]  Shot 5: [1 1 0]
\end{verbatim}

\textbf{Viterbi Decoding:}
\begin{verbatim}
Shot 1: [1 0 1]  Shot 2: [1 0 1]  Shot 3: [0 1 1]
Shot 4: [0 0 1]  Shot 5: [1 1 0]
Confidence: [0.991 0.991 0.991]  [0.991 0.962 0.940]  
            [0.991 0.991 0.991]  [0.991 0.991 0.681]
            [0.991 0.681 0.991]
\end{verbatim}

\textbf{Bayesian Inference:}
\begin{verbatim}
Shot 1: [1 0 1]  Shot 2: [1 0 1]  Shot 3: [0 1 1]
Shot 4: [0 0 0]  Shot 5: [1 1 0]
Confidence: [0.99 0.99 0.99]  [0.99 0.963 0.934]  
            [0.99 0.99 0.981]  [0.99 0.99 0.657]
            [0.989 0.649 0.99]
\end{verbatim}

Notable differences:
\begin{itemize}
\item Shot 4, position 3: Viterbi corrects to $[0,0,1]$ with confidence 0.681, while Majority and Bayesian return $[0,0,0]$ with confidence 0.657. This represents an ambiguous syndrome where Viterbi's global optimization finds a different most-likely error trajectory.
\item Confidence values consistently exceed 0.9 except during transitions, validating the reliability of temporal decoding for stable error configurations.
\end{itemize}


\begin{thebibliography}{99}

\bibitem{shor1995}
P.~W.~Shor, ``Scheme for reducing decoherence in quantum computer memory,'' \textit{Phys. Rev. A} \textbf{52}, R2493 (1995).

\bibitem{steane1996}
A.~M.~Steane, ``Error correcting codes in quantum theory,'' \textit{Phys. Rev. Lett.} \textbf{77}, 793 (1996).

\bibitem{calderbank1996}
A.~R.~Calderbank and P.~W.~Shor, ``Good quantum error-correcting codes exist,'' \textit{Phys. Rev. A} \textbf{54}, 1098 (1996).

\bibitem{gottesman1997}
D.~Gottesman, ``Stabilizer codes and quantum error correction,'' Ph.D. thesis, California Institute of Technology (1997).

\bibitem{shor1996ft}
P.~W.~Shor, ``Fault-tolerant quantum computation,'' \textit{Proc. 37th IEEE Symp. Found. Comp. Sci.}, 56 (1996).

\bibitem{steane1997ancilla}
A.~M.~Steane, ``Active stabilization, quantum computation, and quantum state synthesis,'' \textit{Phys. Rev. Lett.} \textbf{78}, 2252 (1997).

\bibitem{knill1998}
E.~Knill, R.~Laflamme, and W.~H.~Zurek, ``Resilient quantum computation,'' \textit{Science} \textbf{279}, 342 (1998).

\bibitem{aharonov1997}
D.~Aharonov and M.~Ben-Or, ``Fault-tolerant quantum computation with constant error,'' \textit{Proc. 29th ACM Symp. Theory Comput.}, 176 (1997).

\bibitem{preskill1998}
J.~Preskill, ``Fault-tolerant quantum computation,'' \textit{Introduction to Quantum Computation}, World Scientific (1998).

\bibitem{nielsen2000}
M.~A.~Nielsen and I.~L.~Chuang, \textit{Quantum Computation and Quantum Information}, Cambridge University Press (2000).

\bibitem{terhal2015}
B.~M.~Terhal, ``Quantum error correction for quantum memories,'' \textit{Rev. Mod. Phys.} \textbf{87}, 307 (2015).

\bibitem{lidar2013}
D.~A.~Lidar and T.~A.~Brun (eds.), \textit{Quantum Error Correction}, Cambridge University Press (2013).

\bibitem{google2024}
Google Quantum AI, ``Quantum error correction below the surface code threshold,'' \textit{Nature} \textbf{636}, 554 (2024).

\bibitem{quantinuum2024}
C.~Ryan-Anderson et al., ``Implementing fault-tolerant entangling gates on the five-qubit code,'' arXiv:2208.01863 (2022).

\bibitem{ibm2024}
S.~Bravyi et al., ``High-threshold and low-overhead fault-tolerant quantum memory,'' \textit{Nature} \textbf{627}, 778 (2024).

\bibitem{chao2018}
R.~Chao and B.~W.~Reichardt, ``Quantum error correction with only two extra qubits,'' \textit{Phys. Rev. Lett.} \textbf{121}, 050502 (2018).

\bibitem{chamberland2018}
C.~Chamberland and M.~E.~Beverland, ``Flag fault-tolerant error correction with arbitrary distance codes,'' \textit{Quantum} \textbf{2}, 53 (2018).

\bibitem{qiskit2024}
Qiskit Development Team, ``Qiskit: An Open-source Framework for Quantum Computing,'' https://qiskit.org (2024).

\bibitem{fowler2012}
A.~G.~Fowler et al., ``Surface codes: Towards practical large-scale quantum computation,'' \textit{Phys. Rev. A} \textbf{86}, 032324 (2012).

\bibitem{devitt2013}
S.~J.~Devitt, W.~J.~Munro, and K.~Nemoto, ``Quantum error correction for beginners,'' \textit{Rep. Prog. Phys.} \textbf{76}, 076001 (2013).

\bibitem{dennis2002}
E.~Dennis et al., ``Topological quantum memory,'' \textit{J. Math. Phys.} \textbf{43}, 4452 (2002).

\bibitem{aliferis2006}
P.~Aliferis, D.~Gottesman, and J.~Preskill, ``Quantum accuracy threshold for concatenated distance-3 codes,'' \textit{Quantum Inf. Comput.} \textbf{6}, 97 (2006).

\bibitem{tomita2014}
Y.~Tomita and K.~M.~Svore, ``Low-distance surface codes under realistic quantum noise,'' \textit{Phys. Rev. A} \textbf{90}, 062320 (2014).

\bibitem{raussendorf2007}
R.~Raussendorf and J.~Harrington, ``Fault-tolerant quantum computation with high threshold in two dimensions,'' \textit{Phys. Rev. Lett.} \textbf{98}, 190504 (2007).

\end{thebibliography}
\end{document}